\documentstyle[aps,epsfig]{revtex}
\def\Journal#1#2#3#4{{#1} {\bf #2}, #3 (#4)}

\def\PLB{{\em Phys. Lett.}  B}

\def\PRL{\em Phys. Rev. Lett.}

\def\PRD{{\em Phys. Rev.} D}

\begin{document}
\draft
\title{A Supersymmetric Solution to the Solar and Atmospheric
Neutrino Anomalies}
\author{G. Brooijmans}
\address{Institut de Physique Nucl\'eaire, Universit\'e catholique
de Louvain, 1348 Louvain-la-Neuve, Belgium \\
Institut Interuniversitaire des Sciences Nucl\'eaires, Belgium\\
(Now at Fermi National Accelerator Laboratory, P.O. Box 500,
Batavia, Illinois 60510)}
\date{\today}
\maketitle
\begin{abstract}
The formalism for neutrino flavor change induced by lepton family
number violating interactions is developed for
the three-neutrino case, and used to derive the corresponding 
flavor change probabilities in matter.  Applied to the solar and atmospheric
neutrino fluxes, it is argued that the observed anomalies, including the 
zenith dependence for the atmospheric case, could be due to such 
interactions.  
\end{abstract}


\section{Introduction}
\label{intro}

The deficit in the observed solar neutrino fluxes 
(see for example reference \cite{Hat97}), and the deviation
of the atmospheric ``ratio of ratios'' from 1.0 \cite{Fuk98b} are
two of the outstanding problems in today's particle physics.
These are usually interpreted as being due to neutrino 
oscillations, for which, however, there is as yet no 
unambiguous evidence.

In this paper, the possibility that both the solar and atmospheric 
neutrino anomalies could be due to the existence of lepton number
violating interactions, and in particular supersymmetric 
R-parity violating interactions,
is investigated.  Whereas various authors \cite{Dre98,Kon98,Muk98,Chu98}
have recently suggested these
interactions could lead to neutrino masses and mixings leading to
the ``standard'' vacuum oscillation solution, here possible direct,
matter-induced flavor change is studied.

\section{Flavor Changing Interactions and the MSW Effect}
\label{sec:theo}

The time dependency of the flavor composition of 
neutrinos propagating in matter 
is governed by the equation
\begin{equation} 
i \frac{d}{dt}\left(\begin{array}{c} \nu_{e} 
\\ \nu_{\mu} \\ \nu_{\tau} \end{array}\right) =
\left(\begin{array}{ccc} c_{11} + \sqrt{2}\ G_{F} N_{e} & b_{12} & b_{13} \\
b_{12} & c_{22} & b_{23} \\ b_{13} & b_{23} & c_{33} \end{array}\right)
\left(\begin{array}{c} \nu_{e} \\ \nu_{\mu} \\ \nu_{\tau} \end{array}\right),
\label{eq:3numix}
\end{equation}
where the term $\sqrt{2}\ G_{F} N_{e}$ 
-- $N_{e}$ is the electron density --
corresponds to the usual $\nu_{e} - e^{-}$ 
charged-current interactions (this is the term at the 
origin of the classic MSW effect \cite{Wol78,Mik85}), 
and the $c_{ii}$
and $b_{ij}$ coefficients represent flavor-diagonal (FD) 
and flavor-changing (FC)
interactions, respectively.  
Note that we have assumed CP conservation here
(i.e. the propagation matrix is symmetric).
If such interactions exists, they would lead to additional
electroweak potential energy for the propagating neutrino 
eigenstates, thereby opening the possibility
for resonant enhancement of flavor change.
In the Standard Model, electroweak neutral currents are
FD interactions, while FC interactions do not exist in the leptonic sector.

In supersymmetric theories with R-parity violation however, we find
both new FD and FC interactions may exist.  
The two types of interactions that are 
relevant for neutrinos give rise to $\nu - d$-type 
quark and $\nu - e$ scattering via
$d$-type squark and selectron exchange, respectively. 
They correspond to the following two terms in
the lepton number violating part of the superpotential:
\begin{equation}
W = \lambda_{ijk}L_{L}^{i}L_{L}^{j}\overline{E}_{R}^{k} +
\lambda'_{ijk}L_{L}^{i}Q_{L}^{j}\overline{D}_{R}^{k}.
\end{equation}
Here $i,j$ and $k$ are the generation indices, $L_{L}$ and $Q_{L}$ are the 
left-handed lepton and quark doublet superfields, and $E_{R}$ and $D_{R}$
are the right-handed lepton and quark singlets.
Note that the $\lambda$ couplings are antisymmetric in the first 
two indices.

For example,
for neutrino energies small compared to the squark mass
($m_{\tilde{q}}$), the term
\begin{equation}
\frac{\lambda'_{131}\lambda'_{331}}{2m^{2}_{\tilde{b}}}
\left(\overline{\nu_{eL}}\gamma_{\mu}\nu_{\tau L}
\overline{d_{R}}\gamma^{\mu}d_{R}
+ \overline{\nu_{\tau L}}\gamma_{\mu}\nu_{e L}
\overline{d_{R}}\gamma^{\mu}d_{R} \right)
\label{eq:susyex}
\end{equation}
corresponds to flavor-changing neutrino scattering on a $d$-quark
where the $b$-squark acts as the mediator of the interaction.  
We can now rewrite
the coefficients of the propagation matrix in equation 
(\ref{eq:3numix}) in terms
of the R-parity violating couplings.  Taking only into account 
scattering on $d$ quarks
(matter is essentially first-generation) we find for example:
\begin{eqnarray}
b_{12} = \lefteqn{ \left[ \frac{\lambda'_{111}\lambda'_{211}}
{4m^{2}_{\tilde{d}}} + \right.
\frac{\lambda'_{121}\lambda'_{221}}{4m^{2}_{\tilde{s}}} +
\frac{\lambda'_{131}\lambda'_{231}}{4m^{2}_{\tilde{b}}} +
\frac{\lambda'_{112}\lambda'_{212}}{4m^{2}_{\tilde{s}}} +
\frac{\lambda'_{113}\lambda'_{213}}{4m^{2}_{\tilde{b}}}} 
\nonumber \\
& & + \frac{\lambda'_{121}\lambda'_{212}}{4m^{2}_{\tilde{s}}} +
\frac{\lambda'_{131}\lambda'_{213}}{4m^{2}_{\tilde{b}}} +
\frac{\lambda'_{112}\lambda'_{221}}{4m^{2}_{\tilde{s}}} +
\left. \frac{\lambda'_{113}\lambda'_{231}}{4m^{2}_{\tilde{b}}} \right] N_d,
\label{eq:exb12}
\end{eqnarray}
where $N_d$ is the down-quark density in matter.
All the other matrix coefficients may be expressed likewise.

Various authors have suggested \cite{Guz91,Rou91} or established \cite{Bar91}
that the solar neutrino problem could be explained
by the presence of non-standard FD and FC interactions.  The solutions
have recently been analyzed again by
P.I. Krastev and J.N.Bahcall \cite{Kra97} taking into account the newly
available solar neutrino data.

However, all these studies only considered the solar neutrino problem, and
used a simplified two-family model (involving the first and third generations).
In this work, a 3-family analysis is developed, in the attempt to explain both
the solar and atmospheric neutrino anomalies.

The procedure to derive the flavor change probabilities 
in the 3-family case is
as follows.  Let $|\nu_{\alpha}>$ be the flavor 
eigenstates ($\alpha = e,\ \mu$ or
$\tau$) and $|\nu_{i}>$ the propagation eigenstates ($i$=1,2 or 3).  
We then have

\begin{equation}
|\nu_{\alpha}(t)>  =  
\sum_{i=1}^{3} U_{\alpha i} e^{-iHt} |\nu_{i}> 
 =  \sum_{i=1}^{3} U_{\alpha i} e^{-ie_{i}t} |\nu_{i}> 
 =  \sum_{f=1}^{3} \left(\sum_{i=1}^{3} U_{\alpha i} e^{-ie_{i}t}\right) 
U_{fi}^{\dagger} |\nu_{f}>,
\label{eq:nuprop}
\end{equation}
where $U_{\alpha i}$ is a unitary rotation and the $e_{i}$ are the 
propagation energy eigenlevels, 
eigenvalues of the propagation
matrix given in equation (\ref{eq:3numix}).  Since the elements of
the propagation matrix, hence also its eigenvalues, depend on matter density
it is advisable to replace the time $t$ by the distance traveled 
$x$ with the approximation
$t = x$.
The flavor changing amplitude is then given by
\begin{equation}
A\left(\nu_{\alpha}(x=0) \rightarrow \nu_{\beta}(x=L)\right) =
\sum_{i=1}^{3} U_{\alpha i} e^{-ie_{i}L} U_{i \beta}^{\dagger}.
\label{eq:flamp}
\end{equation}
However, since the columns of the rotation matrix $U$ 
which diagonalizes the propagation matrix 
are precisely the eigenvectors of this matrix,one finally obtains
\begin{equation}
P\left(\nu_{\alpha}(x=0) \rightarrow \nu_{\beta}(x=L)\right) =
\left| \sum_{j=1}^{3} V_{j\alpha} e^{-ie_{j}t} V_{j \beta}^{*} \right|^{2},
\label{eq:flprob}
\end{equation}
where $V$ is the matrix whose columns are the eigenvectors of the 
propagation matrix.

\section{Data}
\label{sec:data}

Combining solar and atmospheric neutrino data yields 13 constraints: there
are three different types of solar neutrino experiments, each sensitive to
a different energy range, and the atmospheric ``ratio of ratios'' is 
measured in 5 zenith angle bins for two different energy ranges (sub- and
multi-GeV).

The three types of solar neutrino experiments (Gallium: Sage and Gallex, 
Chlorine:
Homestake, and water \v{C}erenkov: Super-Kamiokande) have different 
lower thresholds
for the neutrino energy.  This has allowed to show that there seems to be an
energy-dependent suppression of the solar neutrino flux (see 
for example N. Hata and
P. Langacker \cite{Hat97}), which could be a strong hint 
for neutrino oscillations,
were it not that neutrinos of different energies are produced 
at different radii
in the Sun and therefore ``see'' different matter densities on their way out.

Table \ref{tab:resdat} summarizes the results from the various experiments
(the results from the Gallium experiments have been combined into one),
and gives the ratio of observed rates over rates predicted by the 
Bahcall-Pinsonneault 95 Standard Solar 
Model \cite{Bah95}.  The experimental results can 
be found in Refs. \cite{Cle96} 
(Homestake), \cite{Abd94} (Sage), \cite{Ham96} (Gallex) 
and \cite{Fuk98a} (Super-Kamiokande).

For the atmospheric neutrino data, the results published recently 
by the Super-Kamiokande Collaboration \cite{Fuk98b} are being used.
Fogli et al. \cite{Fog98} have graphically extracted 
the numerical values from the 
Super-Kamiokande plots, yielding the observed number of data events and 
the expected number of events both for data and Monte-Carlo, such that all the
detection efficiencies can be taken into account.  Table \ref{tab:resdat}
gives the observed ``ratio of ratios'' (ratio of muon-like over 
electron-like events for data divided by the same ratio for Monte-carlo)
for each of the five zenith bins, both for the sub- and multi-GeV samples.
The first error is statistical \cite{Fog98} and the second is 
systematic \cite{Fuk98b}.

We have extracted upper limits on the values of the coefficients $c_{ii}$ and
$b_{ij}$ from the present limits \cite{Roy97,Kim97} on R-parity 
violating couplings.
For reasons of simplicity, only the $\lambda'_{ijk}$ 
couplings (corresponding to
scattering by squark exchange) and scattering on $d$ quarks are included, so 
that the limits are given by equations
analogous to equation (\ref{eq:exb12}).  
These limits have then been transformed to limits on 6 variables
$\epsilon_{11}, \epsilon_{12},\epsilon_{13},\epsilon_{22},
\epsilon_{23}$ and $\epsilon_{33}$ defined by:
\begin{equation}
\epsilon_{ii} = \frac{c_{ii}}{\sqrt{2}\ G_F N_{d}},\
\epsilon_{ij} = \frac{b_{ij}}{\sqrt{2}\ G_F N_{d}},\  i\ne j,
\end{equation}
which give the relative 
strength of the interaction w.r.t. the weak interaction.

Furthermore, we can subtract $c_{11} \times 1$ from the right-hand 
side of equation
(\ref{eq:3numix}).  This does not affect our result since
it is equivalent to a redefinition of the zero of the
neutrino electroweak potential energy.

\section{Analysis and Results}
\label{sec:an+res}

\subsection{Analysis Procedure}

To find the values of the couplings which would 
produce effects similar to those
observed in the data, a $\chi^{2}$-like function with 13 terms
corresponding to the three types of solar neutrino 
experiments and the atmospheric
data is constructed.  The minimum of this $\chi^{2}$ is then searched for
(using Minuit \cite{Min92}),
as a function of the 4 variables
$\epsilon_{13},\epsilon_{23},\epsilon_{22}$ and $\epsilon_{33}$.
$\epsilon_{12}$ has been excluded from the fit since the strong 
limits arising from the non-observation of $\mu\ Ti \rightarrow e\ Ti$
\cite{Roy97} make its contribution to any of the two anomalies negligible.

The Solar model used is
the Bahcall-Pinsonneault Standard Solar Model \cite{Bah95},  
which ``cuts'' the Sun into layers, and for each of these layers, gives
the average electron and neutron 
densities (the Sun is assumed neutral so that the proton density equals 
the electron density) as well as the fraction of neutrinos 
produced for each of the major 
neutrino sources ($pp, \ ^8B, \ ^{13}N,
\ ^{15}O,  ^7Be$ and $pep$).  Our program steps through each layer, 
adds the newly produced neutrinos and changes the flavor of a fraction
of the entering and produced neutrinos in the manner determined 
by the flavor change probability Eq. \ref{eq:flprob}, which is a
function of the variables and the densities.

The Earth is approximated by 5 concentric spheres, each of uniform 
density and composition as has been done 
by Giunti et al. \cite{Giu97}.  To find
the average distance that neutrinos travel through each layer as a function of 
their momentum and zenith angle, a small Monte-Carlo 
simulation has been developed.
The simulation assumes a flat zenith angle distribution in each bin, smears it
with the average angular correlation between the charged lepton and the 
neutrino (55 and 20 degrees for sub-GeV and multi-GeV neutrinos, respectively
\cite{Fuk98b}), and then tracks the neutrinos back through the Earth, computing
the distance traveled through each layer.
The values found are given in Table \ref{tab:avgdis}.

The algebraic expressions for the eigenvectors and eigenvalues of the
propagation matrix as a function of its coefficients 
are fairly complex.  We have used Mathematica \cite{Mat96}
to generate the expression in computer-readable form and 
systematically checked that 
the orthogonality conditions are satisfied.  This is 
done to verify that 
solutions are not produced artificially due to machine accuracy.

\subsection{Results}
\label{subsec:results}

The values of the parameters resulting in the lowest $\chi^{2}$
are given in Table \ref{tab:fitres}.  Table \ref{tab:resdat} 
gives the expected value 
for each of the experimental results for the  best fit parameters.
The minimum $\chi^{2}$ value found is $15.6$, corresponding to a 
probability of $7.6 \%$.

\begin{figure}
\psfig{file=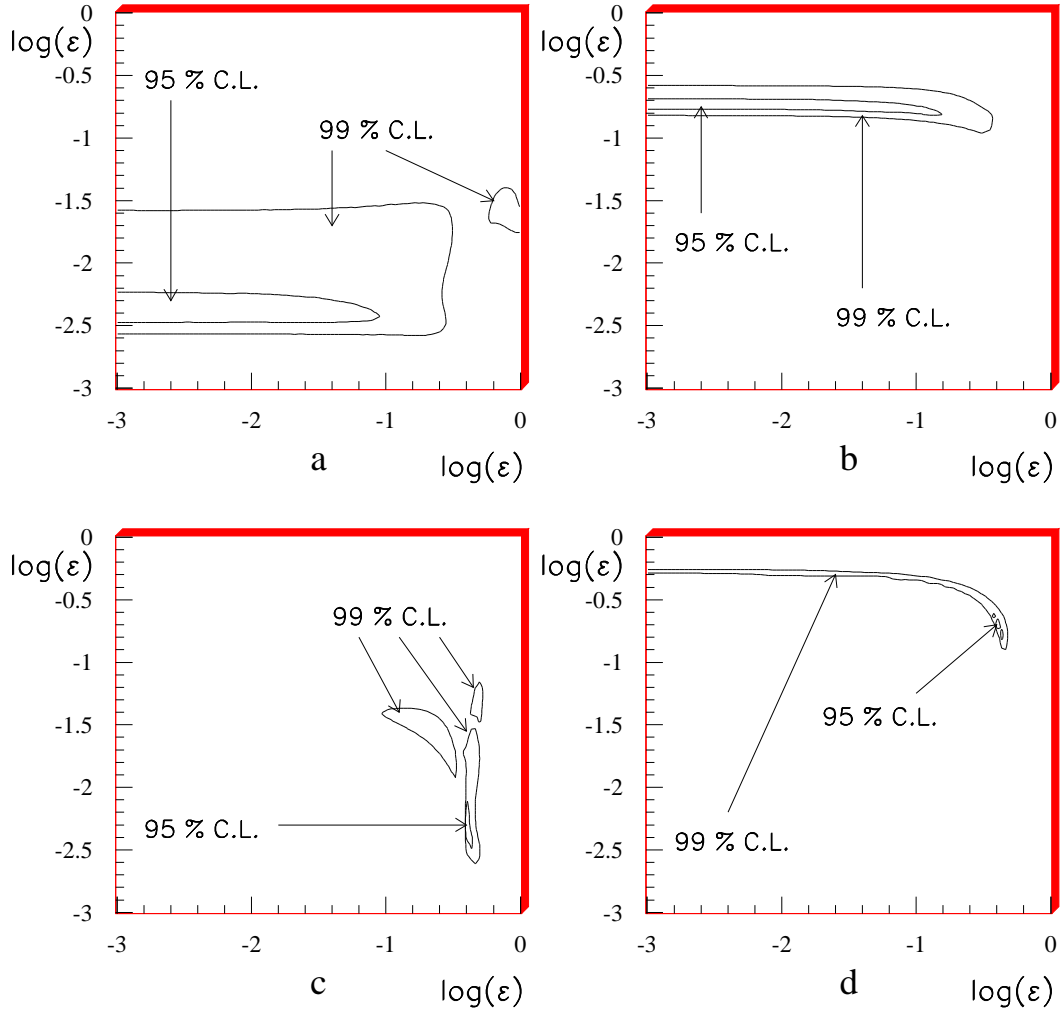,height=15cm}
\caption{\label{fig:excl} Allowed regions in parameter space for
a) $\epsilon_{13}$ versus $\epsilon_{22}$,
b) $\epsilon_{23}$ versus $\epsilon_{22}$,
c) $\epsilon_{13}$ versus $\epsilon_{33}$, and
d) $\epsilon_{23}$ versus $\epsilon_{33}$.
Both 95 and 99 \% C.L. regions are indicated.}
\end{figure}

The variables are somewhat correlated (as expected)
such that changes in their values generally affect the width of the ``good''
region for the others.  Figure \ref{fig:excl} shows the 
region in parameter space allowed at 95 \% and 99 \% C.L. for
each combination of off-diagonal versus diagonal parameter.

Since the value found for $\epsilon_{23}$ has to be significantly larger
than what is allowed by the present non-observation of the decay
$\tau \rightarrow \rho^{0} \mu$, the new interactions must either
only apply to neutrinos, or suffer significant $SU(2)_L$ violation.
It has recently been argued \cite{Berg98}, however, that 
such violation could be in excess of present experimental bounds.

\section{Conclusions}
\label{sec:concl}

We have searched for, and found coupling values of 
a lepton family number violating interaction which would
explain both the solar and atmospheric neutrino anomalies through
an MSW-like 
mechanism in the Sun and in the Earth.
The fit to the data
is not as good as for a purely neutrino oscillation solution, 
and the interaction must either only apply to neutrinos or
undergo significant $SU(2)_L$ violation.  It is however
clear that such an interaction could play a major role in explaining
the anomalies.

It should be noted that in order to have significant flavor change,
neutrinos have to traverse large amounts of matter.  This results in
a large predicted difference between the atmospheric ratio of ratios 
for the first two and last three bins of the Super-Kamiokande multi-GeV
azimuthal dependence histogram.  Increased statistics may make this 
effect visible.

\section{Acknowledgements}

This work was finished while working at Fermi National Accelerator
Laboratory.
The author would like to thank J. Govaerts and G. Gr\'egoire
for useful discussions during this work.


%

\begin{table}
\caption{\label{tab:avgdis} Average distance (in km) traveled by neutrinos 
through each Earth layer (1=core) for each zenith and momentum bin.}
\begin{tabular}{cccccccc}
 Sample &Bin No.&$\cos\theta$& Layer 1& Layer 2& Layer 3& Layer 4& Layer 5 \\
\hline \\
Sub-GeV& 1&$[-1.0,\,-0.6]$ & 245.0 & 1780.8 & 3663.3 & 630.4 & 1131.5 \\
& 2&$[-0.6,\,-0.2]$ & 149.4 & 1130.1 & 2623.4 & 505.4 & 983.7 \\    
& 3&$[-0.2,\,0.2]$ & 81.8  & 649.9 &1731.3 &373.3 &791.1 \\
& 4&$[0.2,\,0.6]$ & 37.6  & 318.5 &1000.0 &243.8 &565.7\\
& 5&$[0.6,\,1.0]$ & 10.9  & 100.9 &396.4 &112.1 &291.6\\    
\hline
Multi-GeV & 1&$[-1.0,\,-0.6]$ & 247.5 & 2148.3 &5157.8 &811.3 &1237.8\\
& 2&$[-0.6,\,-0.2]$ & 8.2   & 236.4 &2398.9 &775.0 &1671.9\\
& 3&$[-0.2,\,0.2]$ & 0.1   & 12.0 & 466.8 &286.5 &1029.4 \\
& 4&$[0.2,\,0.6]$ & 0.0   & 0.2& 34.1 &36.1 &237.7\\
& 5&$[0.6,\,1.0]$ & 0.0   & 0.0 &0.4 &0.8 &11.7\\
\end{tabular}
\end{table}

\begin{table}
\caption{\label{tab:fitres}
Fit variables, maximum allowed absolute value derived 
from the present limits on R-parity violating couplings
(assuming $SU(2)_L$ symmetry), and
lowest $\chi^{2}$ results.  
The corresponding $\chi^{2}$ value is also given.}
\begin{tabular}{lcc}
Variable & Maximum Absolute Value & Best Fit Value \\
\hline
& \\
$\epsilon_{13} $ & $0.421 .  10^{-1}$ & $0.45 . 10^{-2}$ \\
& \\
$\epsilon_{23} $ & $0.708 .  10^{-2}$ & $0.19$ \\
& \\
$\epsilon_{22} $ & $0.164$ & $0.05$ \\
& \\
$\epsilon_{33} $ & $0.460$ & $0.414$ \\ \hline
& \\
$\chi^{2}$ & & $15.6$\\
\end{tabular}
\end{table}

\begin{table}
\caption{\label{tab:resdat} Fit results compared with the data.
The first column indicates each of the 13 constraints and 
the second one gives the observed results: for the solar data the 
fraction of observed rate over the rate predicted by the Bahcall-
Pinsonneault Standard Solar Model 95, and for the atmospheric data
the ``ratio of ratios'' (see text) for each zenith angle bin.
The third column shows the predicted results for the model
described in this paper.}
\begin{tabular}{lcc}
Solar Neutrinos & & \\
Experiment Type & Observed Flux Fraction & Best Fit Flux Fraction\\ 
\hline \\
Homestake & $0.273 \pm 0.021 (exp)\ ^{+0.05}_{-0.03} (theo)$ & 0.368 \\
Gallium & $0.507 \pm 0.049 (exp) \pm 0.028 (theo)$ & 0.424 \\
Super-Kamiokande & $0.366 \ ^{+0.017}_{-0.014} (exp)
\ ^{+0.074}_{-0.046} (theo)$ & 0.372 \\
\hline \\
Atmospheric Neutrinos & & \\
Sub-GeV ($cos\theta$ range) & Observed Ratio of Ratios 
& Best Fit Ratio of Ratios\\ 
\hline \\
$[-1.0,-0.6]$ & $0.41 \pm 0.07 (stat) \pm 0.08 (syst)$ & 0.50 \\
$[-0.6,-0.2]$ & $0.63 \pm 0.07 (stat) \pm 0.08 (syst)$ & 0.50 \\
$[-0.2,0.2]$ & $0.63 \pm 0.07 (stat) \pm 0.08 (syst)$ & 0.58 \\
$[0.2,0.6]$ & $0.82 \pm 0.07 (stat) \pm 0.08 (syst)$ & 0.79 \\
$[0.6,1.0]$ & $0.71 \pm 0.07 (stat) \pm 0.08 (syst)$ & 0.96 \\
\hline \\
Multi-GeV ($cos\theta$ range) & Observed Ratio of Ratios 
& Best Fit Ratio of Ratios\\ 
\hline \\
$[-1.0,-0.6]$ & $0.42 \pm 0.14 (stat) \pm 0.12 (syst)$ & 0.51 \\
$[-0.6,-0.2]$ & $0.52 \pm 0.13 (stat) \pm 0.12 (syst)$ & 0.54 \\
$[-0.2,0.2]$ & $0.70 \pm 0.12 (stat) \pm 0.12 (syst)$ & 0.90 \\
$[0.2,0.6]$ & $0.72 \pm 0.12 (stat) \pm 0.12 (syst)$ & 1.00 \\
$[0.6,1.0]$ & $0.87 \pm 0.16 (stat) \pm 0.12 (syst)$ & 1.00 \\
\end{tabular}
\end{table}

\end{document}